\documentclass[preprint,12pt]{elsarticle}

\usepackage{graphicx,dblfloatfix}
\usepackage{amssymb,amsmath}
\usepackage{lineno}
\usepackage{color}

\newcommand{\blue}{\textcolor{black}}

\journal{Appl. Surf. Sci}
\begin{document}
\begin{frontmatter}

\title{Fast fabrication of optical vortex generators by \blue{femtosecond laser} ablation}

\author{Jian-Guan Hua$^1$, Zhen-Nan Tian$^1$$^*$, Si-Jia Xu$^1$, Stefan Lundgaard$^{2,3}$,\\\protect Saulius Juodkazis$^{2,3}$$^*$} 
\address{$^1$State Key Laboratory of Integrated Optoeletronics, Collee of Electronic Science and Engineering, Jilin University, Changchun 130012, China}
\address{$^2$Centre for Micro-Photonics, Faculty of Science, Engineering and Technology, Swinburne University of Technology, Hawthorn, VIC 3122, Australia}
\address{$^3$Melbourne Centre for Nanofabrication, ANFF, 151 Wellington Road, Clayton, VIC 3168, Australia}
\address{$^*$Email: zhennan\_tian@jlu.edu.cn, saulius.juodkazis@gmail.com }




\begin{abstract}
Fast fabrication of micro-optical elements for generation of optical vortex beams based on the $q$-plate design is demonstrated by femtosecond (fs) laser ablation of gold film on glass. Q-plates with diameter of $\sim$0.5~mm were made in $\sim$1~min using galvanometric scanners with writing speed of 5~mm/s. Period of gratings of $0.8~\mu$m and groove width of 250~nm were achieved using fs-laser ablation at $\lambda = 343$~nm wavelength. Phase and intensity analysis of optical vortex generators was carried out at 633~nm wavelength and confirmed the designed performance. Efficiency of spin-orbital conversion of the $q$-plates \blue{made by ablation of 50-nm-thick film of gold} was $\sim 3\%$. Such gratings can withstand thermal annealing up to 800$^\circ$C. They can be used as optical vortex generators using post-selection of polarisation.  
\end{abstract}

\begin{keyword}
\texttt{nanofabrication\sep ablation\sep q-plates\sep  micro-optics\sep optical vortex}
\end{keyword}

 \end{frontmatter}

\section{Introduction}

Optical vortex generators which produce beams with a spiraling wavefront are used in laser structuring of materials and, recently, for polymerisation~\cite{Ni,Omatsu}. Optical vortices can be \blue{generated} using spatial light modulators (SLMs) with encoded phase patterns upon reflection. \blue{H}owever, for a practical use optical elements for the fixed topological charge, $l$, which defines the azimuthally spiraling phase $e^{il\theta}$ along the beam propagation are sought after for applications where orbital angular momentum (OAM) is \blue{harnessed}.  Miniaturisation of vortex generators is another trend~\cite{09prl103903} which empowers laser tweezer applications for trapping and manipulation of microscopic materials~\cite{trap}. Micro-optical elements for optical vortex generation can be made using spiral plates, which are 3D structures and require 3D printing capability of high precision~\cite{10apl211108}. The spiral plates can reach 100\% efficiency at zero reflection and absorption and transforms light into an optical vortex \blue{independent of the} polarisation state of the incoming light. Among other methods to generate optical vortex, the spin-orbital elements which perform spin, $\sigma$ to orbital, $l$, conversion defines a prominent family. \blue{T}he spin and orbital angular momentum (SAM, OAM) defines the total angular momentum $j=\sigma+l$ by the two independent contributions which can be engineered \blue{independently}. The form-birefringent patterns whose slow-optical axis has a local angular orientation $\theta = q\alpha$ defined by the azimuthal angle $\alpha$, where $q$ is the half-integer, hence, the q-plates~\cite{Biener,Marrucci2006,Sergei}. Q-plates are used to generate beams which carry OAM $l = 2q$ and can be realised using optically anisotropic materials: liquid crystals, semiconductors, silica glass, and different design of metallic or dielectric metasurfaces~\cite{Biener,Shimotsuma,Karimi,Guixin,Jin,Capasso2016,Kruk,Kamali}. \blue{However, one of the most practical solutions of q-plate fabrication by alignment of liquid crystals on a mechanically rubbed surface can only deliver resolution of $\sim 10~\mu$m.}

Direct laser printing of q-plates for \blue{a typical wavelength of $1.5~\mu$m used in telecommunications} and shorter visible wavelengths is a challenging task due to required high aspect ratio grating patterns with duty cycle of 0.5 (grating period  is twice \blue{as large as} the width of the polymer log in the grating)~\cite{17apl181101}. Despite a low efficiency of transmission through nano-grooves made in an optically opaque 50-100~nm thick metal film, the purity (see Sec.~\ref{LG}) of optical vortex generation is very high as defined by portion of the circularly polarised photons which switch their polarisation (spin, $\sigma$) and acquire OAM, $l\neq 0$~\cite{13prl193901,16aom306}. It is of a practical relevance to explore possibility of a fast laser fabrication~\cite{Sun1,Sun2,Xiao,Sun3} of binary $q$-plates by ablation of grooves in metal as compared with more complex material structuring through established nano-lithography and plasma etching for high efficiency and purity optical vortex generators~\cite{18mt}.     
\begin{figure}[tb]
\centering
\includegraphics[width=0.8\textwidth]{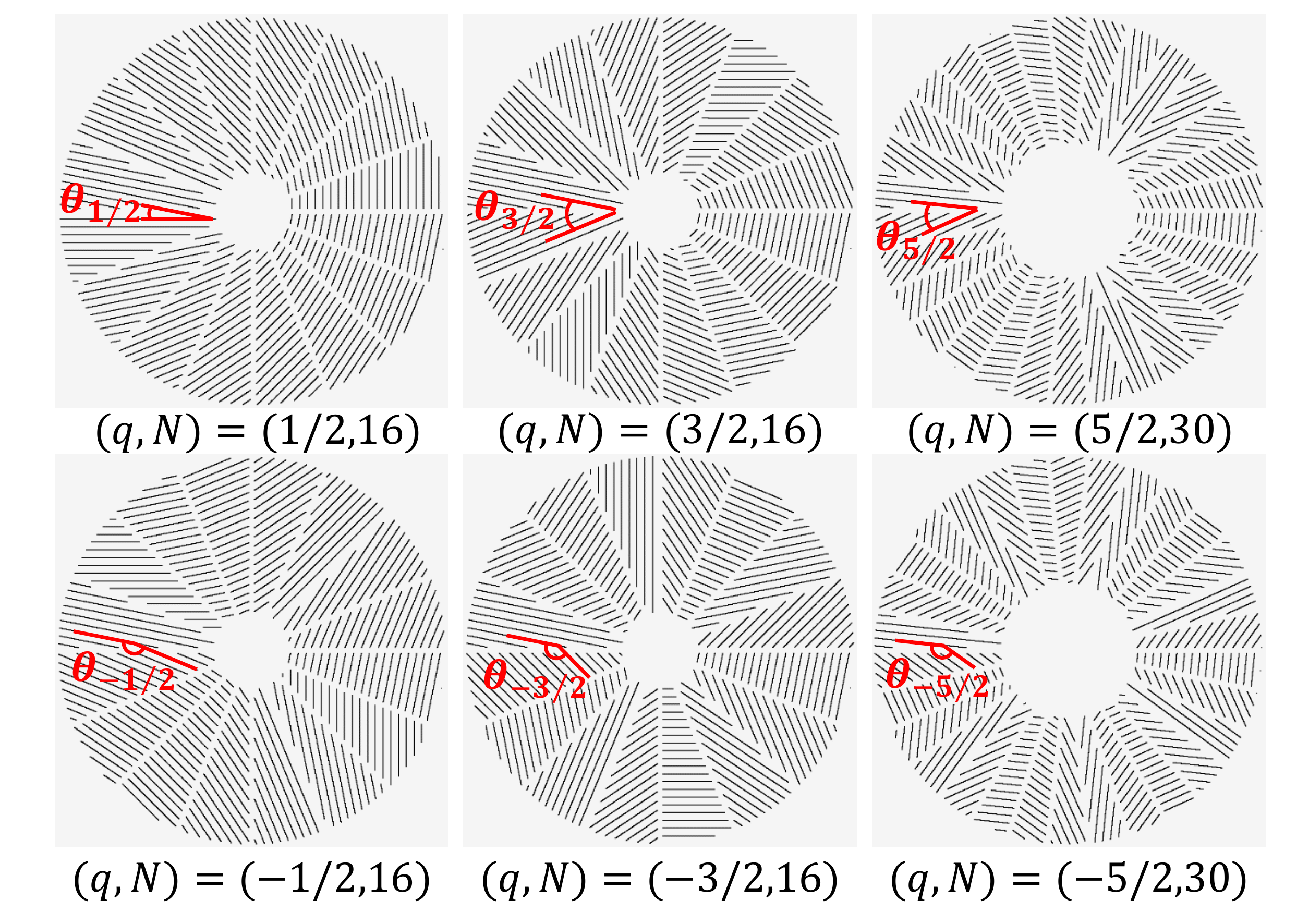}
\caption{Design of q-plates of different $q$ numbers; $q$ values are changing in half-integer steps and can have $\pm$ sign. The number of segments was $N$. These design files with periods of gratings $\Lambda = 0.4; 0.8~\mu$m were prepared for xy-scanners. }\label{f-desi}
\end{figure}
\begin{figure}[tb]
\centering
\includegraphics[width=0.65\textwidth]{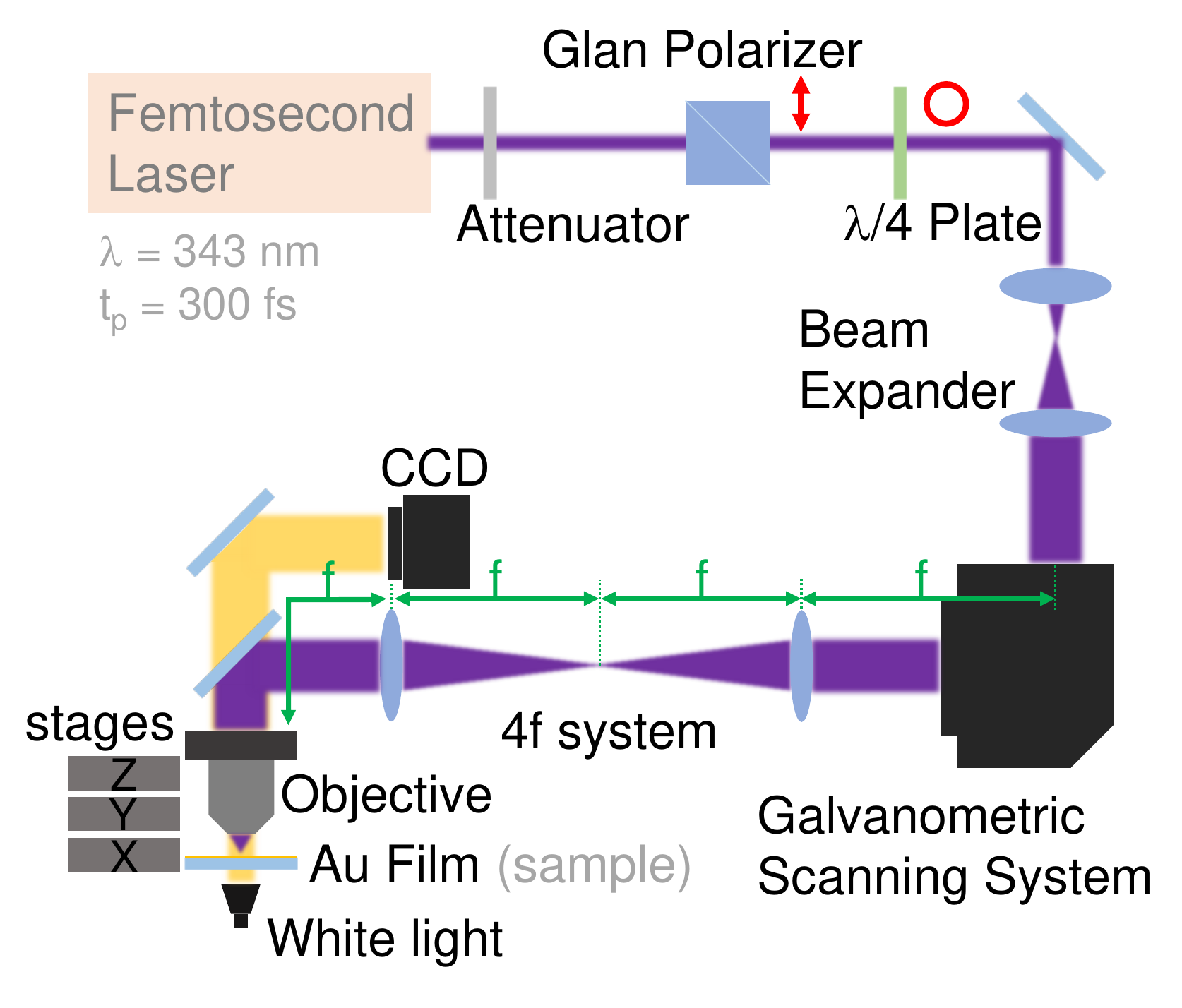}
\caption{ Fs-laser fabrication setup which used \blue{Schwarzschild} reflection objective lens of numerical aperture $NA = 0.5$.}\label{f-setup}
\end{figure}
\begin{figure}[h!]
\centering
\includegraphics[width=0.75\textwidth]{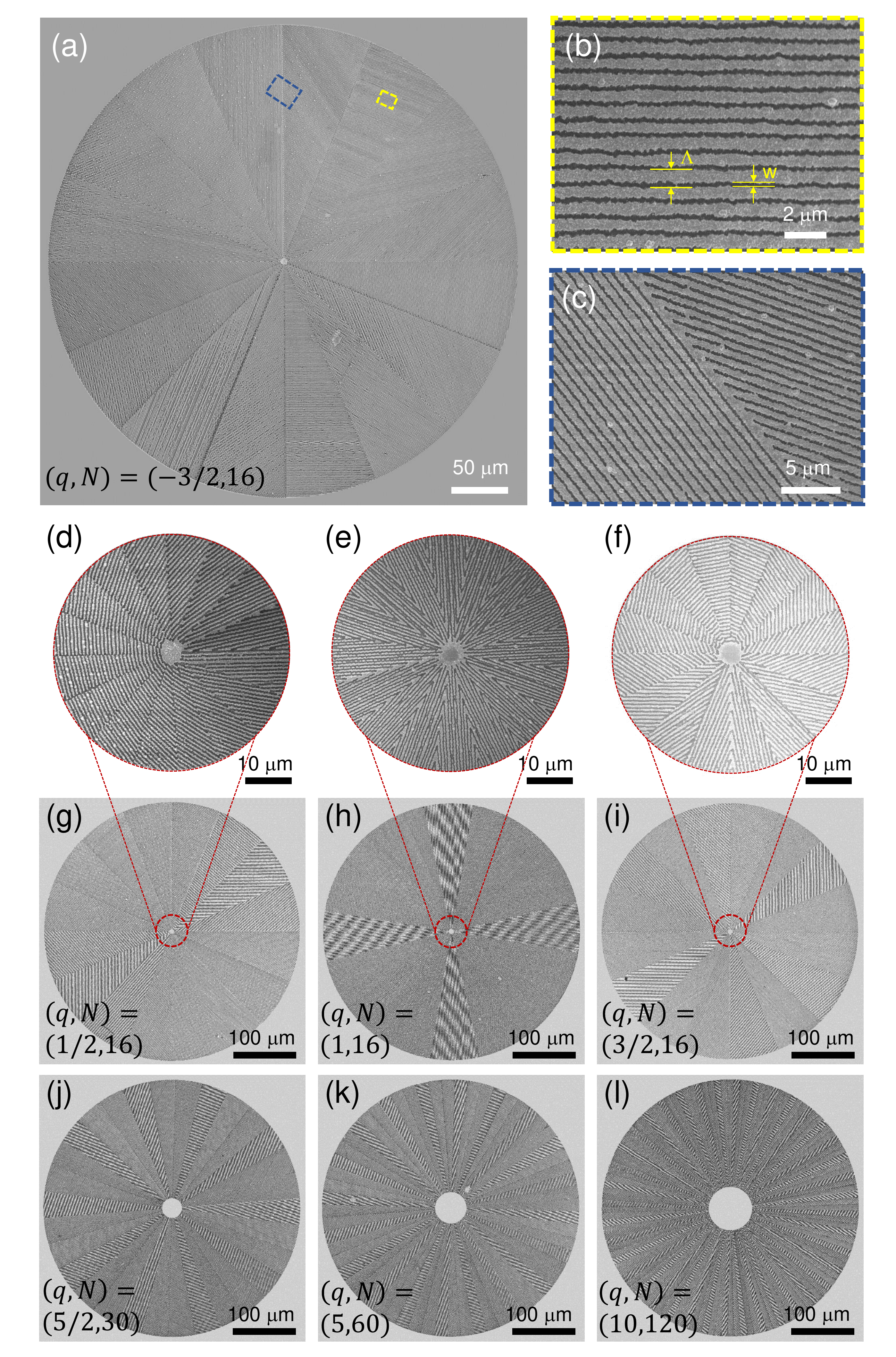}
\caption{(a) SEM image of an $q = -3/2$ plane made of $N = 16$ azimuthally arranged grating segments ablated on 50-nm-thick Au film on glass. (b,c) Magnified SEM images of the two regions marked in (a). The period of grating was $\Lambda = 800$~nm and the ablated line width $w\approx 250$~nm. (d-l) SEM images of ablated $q$-plates with different $q$ values. (a-c) Magnified images of selected $q$-plates. 
 }\label{f-set}
\end{figure}

Here, fabrication of $q$-plates by ablation of gold film is demonstrated with fast writing capability taking only 1~min  per element made on a 0.5~mm diameter area. Phase and intensity patterns of the binary $q$-plates confirmed optical vortex generation at 633~nm wavelength. 

\section{Experimental: samples and procedures}
\subsection{Q-plates: definitions }\label{LG}

The scalar vortex beams are solutions of Helmholtz equation and are defined by the Laguerre-Gaussian (LG) modes~\cite{lg}:
\begin{equation}\label{e3}
E_0^l(r,\theta,z)=\sqrt{\frac{2}{2\pi
l!}}(\sqrt{2}r/w(z))^l(2r^2/w(z)^2)L_0^le^{il\theta}e^{-ikr^2/2q(z)}e^{i\psi},
\end{equation}
where $k=2\pi/\lambda$ is the wavevector,
$\psi(z)=(l+1)\tan^{-1}(z/z_R)$ is the Gouy phase, $q(z)=\pi
R(z)w(z)^2/{\pi w(z)^2-iR(z)\lambda}$ is the complex beam
parameter, $R(z)=z[1+(z_R/z)]$ is the evolving curvature radius,
$w(z)=w_0\sqrt{1+(z/z_R)^2}$ is the evolving beam width, $z_R=\pi
w_0^2/\lambda$ is the Rayleigh lengths, and $L_0^l$ is the
Laguerre polynomials.
 
Q-plates made of grating structures that have an
orientation distribution $\theta(x,y)=q\alpha$, where the
azimuthal angle $\alpha=tan^{-1}(y/x)$ defined in Cartesian
coordinates were fabricated. The emerging electric field,
$\mathbf{E}_{out}$, after passing through the q-plate is
given~\cite{16aom306}:
\begin{equation}\label{e1}
\mathbf{E}_{out}= \mathbf{E}_{in}\tau
[\cos(\Delta/2)\mathbf{e}_\sigma + i\sin(\Delta/2)e^{i2\sigma
q\alpha}\mathbf{e}_{-\sigma}],
\end{equation}
\noindent where $\Delta = \Delta^{'}+i\Delta^{"}$ with
$\Delta^{(',")}=k[n_{\parallel}^{(',")} - n_{\perp}^{(',")}]h$ are
defining the retardance $\Delta^{'}$ and dichroism $\Delta^{"}$
for the $h$ height of q-plate (along the light propagation length)
at wavevector $k = 2\pi/\lambda$ for wavelength $\lambda$, and $\tau = e^{-k(n_{\parallel}^{"} + n_{\perp}^{"})h/2}$ is
transmittance which accounts for losses and defines the efficiency
of the spin-orbital coupling. The incident field is circularly
polarised $\mathbf{E}_{in}\propto \mathbf{e}_{\sigma = \pm1}$ with
$\mathbf{e}_{\sigma} = \frac{1}{\sqrt{2}}(\mathbf{x} +
i\sigma \mathbf{y})$ and $\sigma = \pm 1$ defining the left and
right-handed polarisations in the Cartesian frame, respectively.
The spin-orbital coupling efficiency is determined by the
dichroism and birefringence properties of the material and is
maximised for the half-wavelplate condition of the q-plate:
$\Delta^{'} = \pi$ modulo $2\pi$.

The purity of optical vortex generation $\eta =
|\mathbf{E}_{out}\cdot \mathbf{e}_\sigma|^2/|\mathbf{E}_{out}|^2 =
\sin^2(\Delta'/2)$ defines the relative weight of the circularly
polarised output field component carrying the optical
vortex~\cite{16aom306}. When optical losses are present due to
dichroism, the purity is given as~\cite{DavitPhD}:
\begin{equation}\label{e2}
\eta = \frac{1}{2}\left[ 1 - \frac{\cos\Delta'}{\cosh\Delta"}
\right].
\end{equation}
\noindent At the half-waveplate condition $\Delta^{'} = \pi$,
$\eta = 1$ when dichroic losses are absent $\Delta^{"} = 0$.
Dichroic losses are measured for the $q=0$ plate (a grating) for
two linearly polarised beam orientations are $\tau \equiv
e^{-\Delta^{"}} = \sqrt{P_\parallel/P_\perp}$. This defines limitation for the achievable purity from the material point of view. Experimentally measured purity is the power of the optical vortex component with $l\neq 0$ to the total transmitted power through the $q$-plate (the total transmission through the 50-nm-thick Au film was smaller than 0.8\% at all visible wavelengths, see, Sec.~\ref{suppl}). 

Thin binary transmission masks have weak birefringence~\cite{Iwami} and low transmission, hence, the efficiency of vortex generation are expected to be low.  The retardance is defined by a phase shift between TE and TM modes through the thin film (50~nm in our case)~\cite{Iwami}. 
The pair $(\Delta^{'},\Delta^{"})$ is defining the purity (Eqn.~\ref{e2}). 
The dichroic ratio used to characterise grid polarisers $D_r=A_\parallel/A_\perp$ defined by the absorbance of E-field $\parallel$ to the grid (high $A$) and that perpendicular $\perp$ (low $A$) is directly related to the dichroic losses $\Delta^{"}$. When $D_r = 1$ there is no anisotropy of absorbance (also transmittance, $T$).

Q-plate patterns were designed for \blue{control of a} laser fabrication software (Fig.~\ref{f-desi}). The number of segments, $N$ were maximised to obtain the most efficient addition of light emerging through the grating slits. For the large $q$ values the central part of the element was not structured, when there was less than one period fitting into \blue{the} segment \blue{section} at small radius. For the sake of fabrication flexibility, $q$-plates with different $q$ signs $\pm$ were fabricated (Fig.~\ref{f-desi}). For the actual applications, the sign of vortex beam (OAM $l$) can be changed by changing the sign of the incoming SAM $\sigma$.     

\begin{figure}[tb]
\centering
\includegraphics[width=0.7\textwidth]{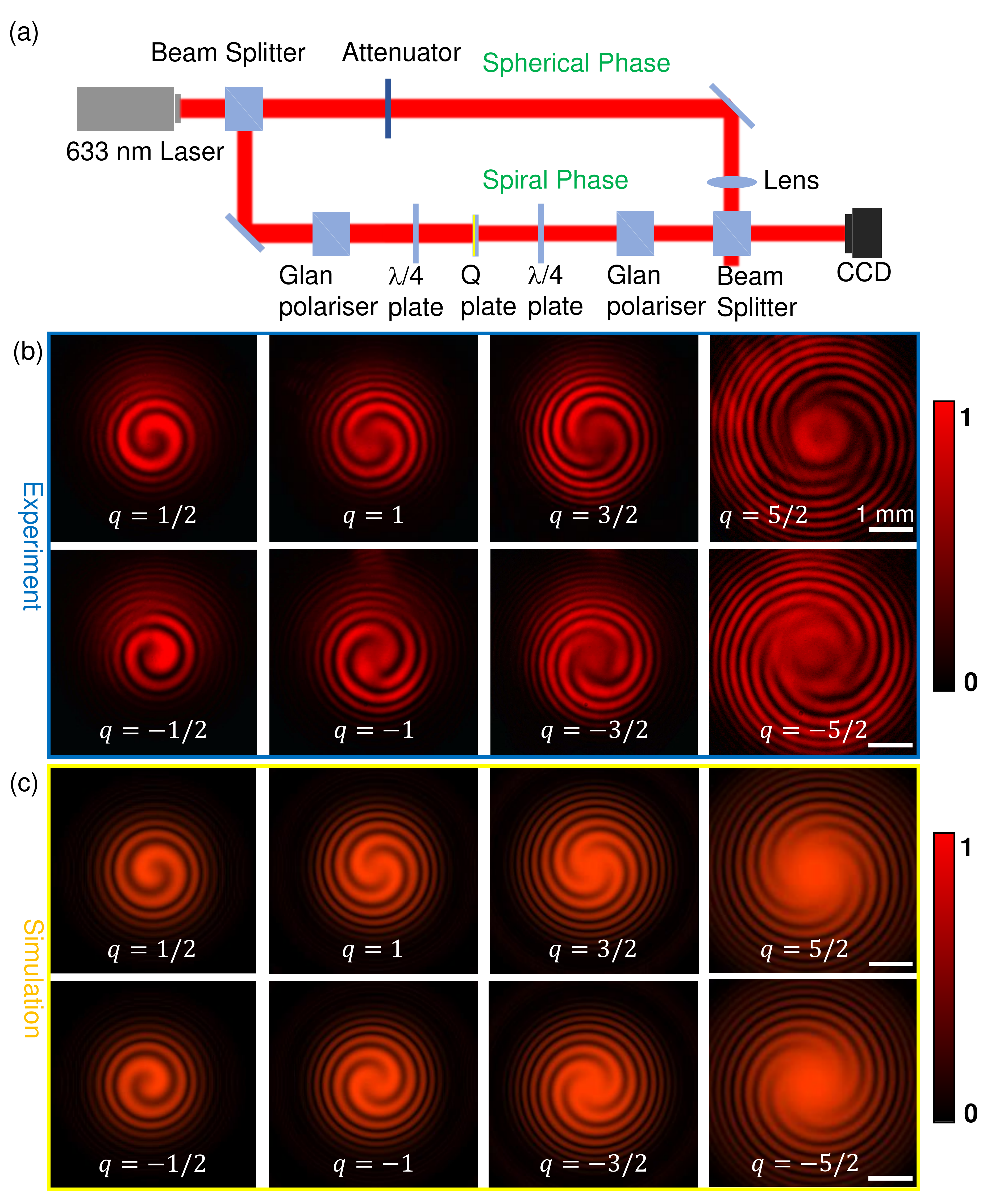}
\caption{(a) A phase characterisation setup of the optical vortex beams generated by $q$-plates. (b) Experimental interferograms of different $q$-plates. \blue{Scale bars 1~mm.} (c) Numerical simulations of the phase of the corresponding LG vortex beams (Eqn.~\ref{e3}).       
 }\label{f-images}
\end{figure}

\subsection{Fabrication and characterisation}
 
Q-plates were \blue{fabricated by ablation of a 50-nm-thick gold film deposited} on \blue{fused} silica substrates \blue{(Jinlong photoelectric, Ltd.)} by magnetron sputtering; a few nanometers Cr adhesion layer was used between Au and silica. Figure~\ref{f-setup} shows setup used for laser fabrication. It is based on \blue{a Yb:KGW solid state} fs-laser (Pharos, Light Conversion Ltd.), galvo-scanners (Sunny Technology Ltd.), and reflection-type \blue{Schwarzschild} objective lens of numerical aperture $NA = 0.5$ (Thorlabs, LMM-40X-UVV). 

A high 200~kHz repetition rate fs-laser at $\lambda = 343$~nm wavelength was obtained via an internally integrated triple frequency system \blue{(Pharos)}. The shorter wavelength was used to achieve higher machining resolution. The high repetition enables higher processing speeds and improves processing efficiency. The pulse width of the fs-laser was $t_p = 300$~fs. First, the laser passes through the Glan prism and the $\lambda/4$ plate to generate a circular polarization in order to avoid the influence of polarization under high numerical aperture focusing on the spot shape and reduce effects related to directionality of electronic heat diffusion depending on the scanning direction~\cite{16aom1209}. The laser then passes through a $3^\times$ beam expander system to obtain a larger spot size and is guided through the galvanometric xy-scanners, which can \blue{deliver a} high scanning speed of 5~mm/s \blue{(on sample)} used in this study. Rapid movement of the focal spot was key to processing efficiency. \blue{An i}nternal laser shutter was used for blocking the beam during a change of location required by the drawing algorithm. The 4f optical system consisting of a double lens was implemented for \blue{transferring} of the deflection angle of the galvanometric scanners onto the objective entrance. We also used a high numerical aperture UV transmissive objective (Nikon, $20^\times$ magnification, $NA=0.75$) which had 60\% transmittance at the 343~nm wavelength. The laser power measured before the entrance to the objective lens was 1.1~mW (or pulse energy $E_p = 3.3$~nJ/pulse on the sample).

Interestingly, the \blue{Schwarzschild} objective lens was delivering a better quality lines at the end points. \blue{The diameter of the focal spot was $d = 1.22\lambda/NA = 837$~nm. Ablation was carried out at $E_p = 4$~nJ/pulse (at objective transmission of $T = 33\%$ at $\lambda=343$~nm), at repetition rate of 200~kHz and at $NA = 0.5$ focusing. This corresponds to the fluence per pulse of $F_p = E_p/[\pi(d/2)^2]=0.24$~J/cm$^2$ and is close to the threshold of ablation of metals and the average irradiance was $I_p = F_p/t_p = 0.8$~TW/cm$^2$.  The laser ablation features smaller than the diffraction limit at the used focusing $\frac{d}{2} = 418$~nm can be printed on the surface by ablation due to the threshold effect as it was the case in this study.} Laser ablated patterns were characterised using scanning electron microscopy (SEM, Jeol JSM-7500F).

For numerical modeling of the phase and intensity of the vortex beams we used Virtual Lab Fusion (LightTrans) software.

\begin{figure}[tb]
\centering
\includegraphics[width=0.95\textwidth]{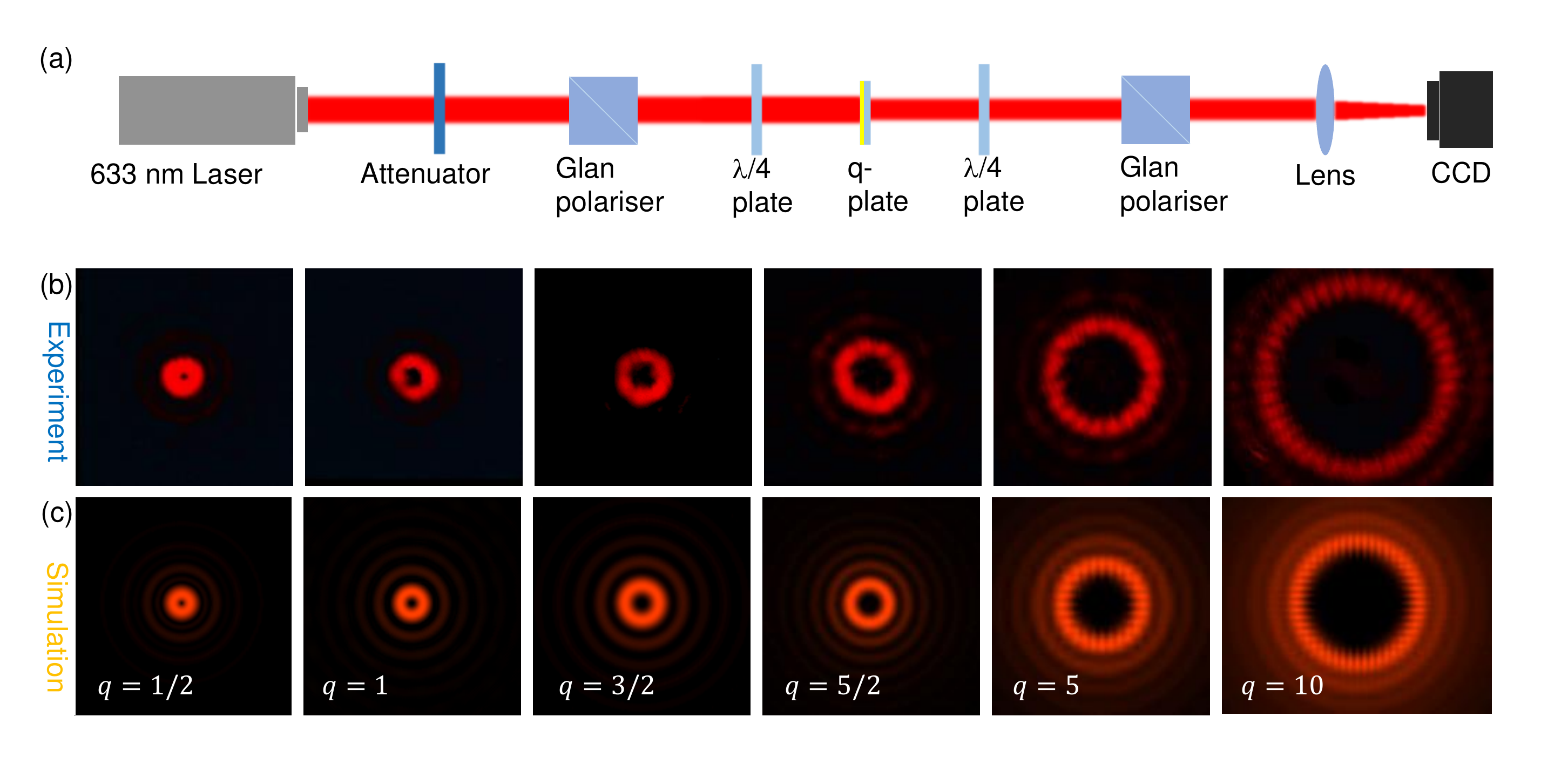}
\caption{(a) Setup for intensity readout of the optical vortex beams generated by $q$-plates. (b) Experimental readout of intensity from different $q$-plates. (c) Numerical simulations of the intensity of the corresponding LG vortex beams (Eqn.~\ref{e3}).       
 }\label{f-inten}
\end{figure}
\begin{figure}[tb]
\centering
\includegraphics[width=0.8\textwidth]{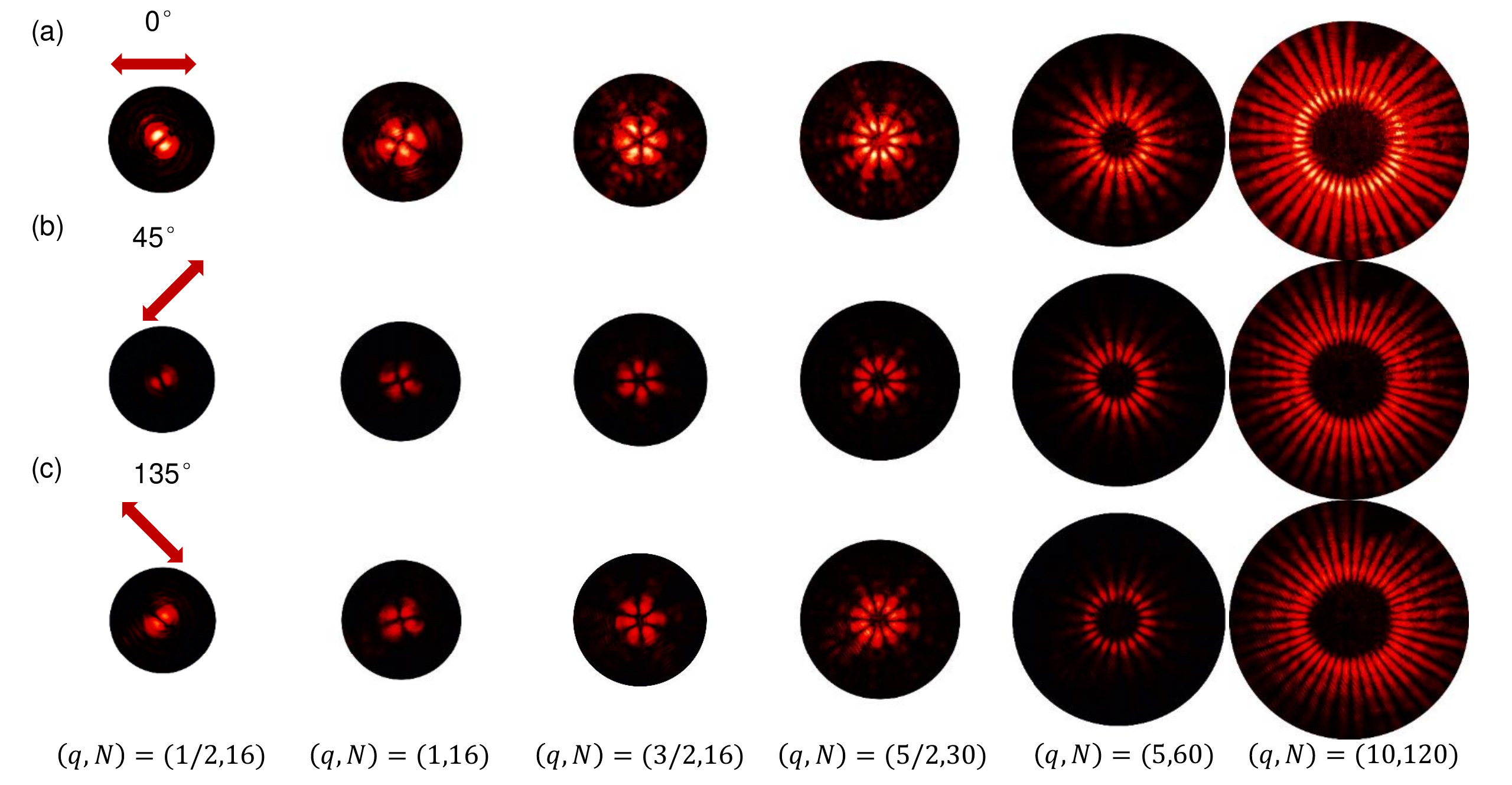}
\caption{Intensity readout of different $q$-plates with linearly polarised beam at: $0^\circ$ (a), $45^\circ$ (b) and $135^\circ$ (c). The segment of $q$-plate which is rotated at $45^\circ$ to the incident polarisation is contributing the largest intensity.} \label{f-segm}
\end{figure}

\section{Results and Discussion}
      
Laser ablated pattern of $q=-3/2$ plate  composed of $N = 16$ grating segments is shown in Fig.~\ref{f-set}. The period $\Lambda = 0.8~\mu$m with opening of groove $w\approx 250$~nm was fabricated in 43~s. The diameter of the $q$-plate was 0.4~mm. Close proximity of the segments with separation $\sim 1~\mu$m was achieved with high fidelity (Fig.~\ref{f-set})(c)) without widening of the lines at the end point of a scan lines. Even periods of $\Lambda = 0.4~\mu$m were achieved with a similar ablated groove width. Flexibility of this approach is summarised in Fig.~\ref{f-set}\blue{(d-l)} showing different patterns reaching \blue{the} demanding $q =10$ and $N=120$ conditions. The most complex $q$-plates attempted took only approximately $2.5^\times$ longer as compared with the simplest ones. \blue{The same resolution of fabrication was achieved for the most complex patterns at the writing/printing speed used in this study.}      
      
To check the property of optical vortex generation, the spiraling phase was revealed using two arms interferometer (Fig.~\ref{f-images}(a)). The used 633~nm wavelength had only 0.22\% transmission through the unstructured Au film (Sec.~\ref{suppl}, Fig.~\ref{f-50nm}). A good \blue{qualitative} match between the experimental interferograms and calculated by Virtual Lab Fusion (LightTrans) was corroborated. \blue{The vortex beam size and phase structure were following that expected from design parameters of $q$-plate.}  

\blue{For optical characterisation, polarisation state was controlled in the interferometer.} The second $\lambda/4$ plate \blue{(Fig.~\ref{f-images}(a))} was set at $\pi/2$ to the first one (before the q-plate). With this setting, the circularly polarised beam with $+\sigma$ and $l = 0$ (a plane wave before the q-plate) becomes linearly polarised and the second Glan polariser is at the crossed orientation to \blue{the incoming} polarisation. The vortex beam generated via spin-orbital coupling has emerging photons with $l\neq 0$ with opposite spin $-\sigma$ which are transmitted through the second Glan polariser. In the reference arm, an attenuator and a lens were added to obtain the best contrast of the interferogram of the two  beams with spherical wavefronts. \blue{A q}ualitative match with \blue{the} experimentally observed interference patterns was obtained numerically using Laguerre-Gauss (LG) beams \blue{(Eqn.~\ref{e3})} interference with a Gaussian intensity envelope (Fig.~\ref{f-images}(c)).   

The efficiency of the spin-orbital conversion was tested by measuring power after the second Glan polariser at two orthogonal orientations of the low and high transmittance and the value of 3.1\% was determined as power of the $l\neq 0$ mode to the total power~\cite{13prl193901} (see, Sec.~\ref{f-D} for determination of optical slit retardance which defines $\Delta^{'}$). It is comparable with purely dielectric polymerised $q$-plates~\cite{17apl181101}. All dielectric $q$-plates with duty cycle of 0.5 can potentially reach high efficiency and purity (Eqns.~\ref{e1},~\ref{e2}) of optical vortex generation, but it is very demanding to satisfy the required effective structure height for the $\pi$ phase retardance for an effectively lower mass density pattern. Moreover, the laser printing throughput for sub-1~mm areas will take up to $10^3$ times longer even for $\sim 1~\mu$m-tall $q$-plates \blue{as in ref.~\cite{17apl181101}}. 

To determine the dichroic ratio $D_r$ a grating ($q=0$ plate) was fabricated over the 0.4-mm-diameter area and transmission of a linearly polarised light  perpendicular ($T_{max}$) and parallel ($T_{min}$) to the grating wires was measured $D_r = \lg(T_{min})/\lg(T_{max}) = 1.32$; for a grating  $D_r>1$ it is expected. The dichroism estimated \blue{f}rom ratio of transmitted powers was $\Delta^{"} = 0.04$. Transmission through a $q=1/2$ plate was measured using a reference, an ablated 0.4-mm-diameter hole in Au, and was 76.6\%. A slightly lower transmission is expected for larger $q$ values due to increasing central region of unstructured $q$-plate. 

\blue{Thermal stability of a high resolution $q$-plate pattern was tested next.} High temperature annealing at 830$^\circ$C for 40~min was able to reduce random pattern of debris nanoparticles generated by ablation (See Sec.~\ref{suppl}). This \blue{marginally} improved transmission \blue{and would need a dedicated study of thermal treatment effects on different thickness gold films, grating period and duty cycle}. Importantly, \blue{the thermal test} also shows that the fabricated $q$-plates can withstand high irradiance conditions and will not deteriorate in the high power laser beam due to a comparatively lower transmission compared with dielectric $q$-plates. Good adhesion and thermal stability of Au film on silica can be harnessed for a hard mask in Ar plasma etching. This can increase the retardance of the structure linked to $\Delta^{'}$ and the purity of the optical vortex generation and will be investigated in future.       
      
\blue{Geometry of the generated vortex beams and uniformity of fabricated segments were further investigated by direct imaging of the intensity distribution.} The intensity readout of the optical vortex beam shows the doughnut shape (Fig.~\ref{f-inten}) with the peak intensity position defined by the radius $r = w_0\sqrt{l/2}$, where $w_0$ is the waist of the Gaussian TEM$_{00}$ mode. Using \blue{a} circularly polarised illumination of the $q$-plate \blue{the intensity distribution after a compensating $\lambda/4$ plate is shown in (b) for the $l\neq 0$ vortex beam portion}. 

The fabricated $q$-plates are \blue{composed} out of grating segments and their position in the optical element can be visualised using linearly polarised light illumination as shown in Fig.~\ref{f-segm}. \blue{When orientation of the grating was at $\pi/4$ with respect to the linear polarisation of the illuminating beam, the strongest intensity was observed due to the form birefringence. The number of bright lobes was $4q$; a typical Maltese cross like pattern is observed for the radial orientation pattern of $q=1$. Uniformity of side lobes is testimonial for the same ablation pattern quality among the segments of $q$-plate. }         
      
\section{Conclusions}

Fast $\sim$1~min fabrication of $\sim 0.5$~mm diameter $q$-plates of different complexity is demonstrated with resolution of 250~nm grooves and period of 800~nm on gold coated silica. This simple binary grating design delivered a 3.1\% efficiency of the spin-to-orbital conversion. The quality of the optical doughnut beam intensity and phase patterns were corroborated using interferometry. By using post-selection polarisation optics \blue{(as used in beam characterisation in this study),}  vortex generators can be \blue{delivered}. They \blue{could} withstand high laser powers and light induced heating as \blue{can be inferred from the} thermal treatment of the $q$-plates. Linear scan speeds were reaching 5~mm/s with pulse to pulse shift by $25$~nm at 200~kHz repetition rate (or 33.5 pulses per $1.22\lambda/NA = 0.84~\mu$m diameter focal spot) and ablated 250-nm-wide lines in a 50-nm-thick gold film. As it was established earlier, such optical vortex generators have optical broadband ``white'' performance~\cite{16aom306}. \blue{A more efficient vortex beam generation and a higher purity of spin-orbital conversion can be envisaged after a plasma etching using the ablated $q$-plate pattern as a hard mask.}    

\section*{References}
\bibliographystyle{elsarticle-num}

\begin{thebibliography}{10}
\expandafter\ifx\csname url\endcsname\relax
  \def\url#1{\texttt{#1}}\fi
\expandafter\ifx\csname urlprefix\endcsname\relax\def\urlprefix{URL }\fi
\expandafter\ifx\csname href\endcsname\relax
  \def\href#1#2{#2} \def\path#1{#1}\fi

\bibitem{Ni}
J.~Ni, C.~Wang, C.~Zhang, Y.~Hu, L.~Yang, Z.~Lao, B.~Xu, J.~Li, D.~Wu, J.~Chu,
  Three-dimensional chiral microstructures fabricated by structured optical
  vortices in isotropic material, Light: Sci. Appl. 6 (2017) e17011.

\bibitem{09prl103903}
E.~Brasselet, N.~Murazawa, H.~Misawa, S.~Juodkazis, Optical vortices from
  liquid crystal droplets, Phys. Rev. Lett. 103 (2009) 103903.

\bibitem{10apl211108}
E.~Brasselet, M.~Malinauskas, A.~\v{Z}ukauskas, S.~Juodkazis, Photo-polymerized
  microscopic vortex beam generators : precise delivery of optical orbital
  angular momentum, Appl. Phys. Lett. 97 (2010) 211108.

\bibitem{Biener}
G.~Biener, A.~Niv, V.~Kleiner, E.~Hasman, Formation of helical beams by use of
  {Pancharatnam - Berry} phase optical elements, Opt. Lett. 27 (2002) 1875 --
  1877.

\bibitem{Marrucci2006}
L.~Marrucci, C.~Manzo, D.~Paparo,
  \href{http://link.aps.org/doi/10.1103/PhysRevLett.96.163905}{Optical
  spin-to-orbital angular momentum conversion in inhomogeneous anisotropic
  media}, Phys. Rev. Lett. 96 (2006) 163905.
\newblock \href {http://dx.doi.org/10.1103/PhysRevLett.96.163905}
  {\path{doi:10.1103/PhysRevLett.96.163905}}.
\newline\urlprefix\url{http://link.aps.org/doi/10.1103/PhysRevLett.96.163905}

\bibitem{Sergei}
S.~Slussarenko, A.~Murauski, T.~Du, V.~Chigrinov, L.~Marrucci, E.~Santamato,
  \href{http://www.opticsexpress.org/abstract.cfm?URI=oe-19-5-4085}{Tunable
  liquid crystal q-plates with arbitrary topological charge}, Opt. Express
  19~(5) (2011) 4085--4090.
\newblock \href {http://dx.doi.org/10.1364/OE.19.004085}
  {\path{doi:10.1364/OE.19.004085}}.
\newline\urlprefix\url{http://www.opticsexpress.org/abstract.cfm?URI=oe-19-5-4085}

\bibitem{Shimotsuma}
Y.~Shimotsuma, P.~G. Kazansky, J.~Qiu, K.~Hirao, Self-organized nanogratings in
  glass irradiated by ultrashort light pulses, Phys. Rev. Lett. 91 (2003)
  247405.

\bibitem{Karimi}
E.~Karimi, S.~A. Schulz, I.~De~Leon, H.~Qassim, J.~Upham, R.~W. Boyd,
  \href{http://search.proquest.com/docview/1793427063?accountid=14205}{Generating
  optical orbital angular momentum at visible wavelengths using a plasmonic
  metasurface}, Light Sci Appl 3~(5) (2014) 4.
\newline\urlprefix\url{http://search.proquest.com/docview/1793427063?accountid=14205}

\bibitem{Guixin}
G.~Li, M.~Kang, S.~Chen, S.~Zhang, E.~Y.-B. Pun, K.~W. Cheah, J.~Li,
  \href{http://dx.doi.org/10.1021/nl401734r}{Spin-enabled plasmonic
  metasurfaces for manipulating orbital angular momentum of light}, Nano Lett
  13~(9) (2013) 4148--4151.
\newblock \href {http://arxiv.org/abs/http://dx.doi.org/10.1021/nl401734r}
  {\path{arXiv:http://dx.doi.org/10.1021/nl401734r}}, \href
  {http://dx.doi.org/10.1021/nl401734r} {\path{doi:10.1021/nl401734r}}.
\newline\urlprefix\url{http://dx.doi.org/10.1021/nl401734r}

\bibitem{Jin}
J.~Jin, J.~Luo, X.~Zhang, H.~Gao, X.~Li, M.~Pu, P.~Gao, Z.~Zhao, X.~Luo,
  Generation and detection of orbital angular momentum via metasurface, Sci.
  Reports 6 (2016) 24286.

\bibitem{Capasso2016}
R.~C. Devlin, A.~Ambrosio, D.~Wintz, S.~L. Oscurato, Z.~A. Yutong,
  M.~Khorasaninejad, J.~Oh, P.~Maddalena, F.~Capasso, Spin-to-orbital angular
  momentum conversion in dielectric metasurfaces, e-prints ArXiv :1605.03899
  (2016) arXiv\href {http://arxiv.org/abs/http://arxiv.org/abs/1605.03899}
  {\path{arXiv:http://arxiv.org/abs/1605.03899}}.

\bibitem{Kruk}
S.~Kruk, B.~Hopkins, I.~I. Kravchenko, A.~Miroshnichenko, D.~N. Neshev, Y.~S.
  Kivshar, nvited article: Broadband highly efficient dielectric metadevices
  for polarization control, Appl. Phys. Lett.: Photonics 1 (2016) 030801.

\bibitem{Kamali}
S.~M. Kamali, E.~Arbabi, A.~Arbabi, Y.~Horie, A.~Faraon, Highly tunable elastic
  dielectric metasurface lenses, Laser Photn. Rev. 10 (2016) DOI:
  10.1002/lpor.201600144.

\bibitem{17apl181101}
X.~Wang, A.~A. Kuchmizhak, E.~Brasselet, S.~Juodkazis, Dielectric geometric
  phase optical elements fabricated by femtosecond direct laser writing in
  photoresists, Appl. Phys. Lett. 110~(18) (2017) 181101.

\bibitem{13prl193901}
E.~Brasselet, G.~Gervinskas, G.~Seniutinas, S.~Juodkazis, Topological shaping
  of light by closed-path nanoslits, Phys. Rev. Lett. 111~(19) (2013) 193901.

\bibitem{16aom306}
D.~Hakobyan, H.~Magallanes, G.~Seniutinas, S.~Juodkazis, E.~Brasselet,
  Tailoring orbital angular momentum of light in the visible domain with
  metallic metasurfaces, Adv. Opt. Mater. 4 (2016) 306 -- 312.

\bibitem{Sun1}
Y.-L. Zhang, Q.-D. Chen, H.~Xia, H.-B. Sun, Designable {3D} nanofabrication by
  femtosecond laser direct writing, Nano Today 5 (2010) 435 -- 448.

\bibitem{Sun2}
H.~Xia, J.~Wang, Y.~Tian, Q.-D. C. X.-B. Du, Y.-L. Zhang, Y.~He, H.-B. Sun,
  Ferrofl uids for fabrication of remotely controllable micro-nanomachines by
  two-photon polymerization, Adv. Mat. (22) 3204 -- 3207.

\bibitem{Xiao}
Y.~Zhang, L.~Guo, S.~Wei, Y.~Hec, H.~Xia, Q.-D. Chen, H.-B. Sun, F.-S. Xiao,
  Direct imprinting of microcircuits on graphene oxides film by femtosecond
  laser reduction, Nano Today 5 (2010) 15 -- 20.

\bibitem{Sun3}
L.~Wang, Q.-D. Chen, X.-W. Cao, R.~Buividas, X.~Wang, S.~Juodkazis, H.-B. Sun,
  Plasmonic nano-printing: large-area nanoscale energy deposition for efficient
  surface texturing, Light: Sci. Appl. 6 (2017) e17112.

\bibitem{18mt}
G.~Seniutinas, E.~Brasselet, A.~Balcytis, C.~David, S.~Juodkazis, Diamond: a
  gem for micro-optics, Materials Today (2018) (in press).

\bibitem{lg}
L.~Allen, M.~W. Beijersbergen, R.~J.~C. Spreeuw, J.~P. Woerdman, Orbital
  angular momentum of light and the transformation of {Laguerre-Gaussian} laser
  modes, Phys. Rev. A 45~(11) (1992) 8185 -- 8189.

\bibitem{DavitPhD}
D.~Hakobyan, Spin-orbit optomechanics of space-variant birefringent media,
  Ph.D. thesis, University of Bordeaux and Swinburne University of Technology
  (2016).

\bibitem{Iwami}
K.~Iwami, M.~Ishii, Y.~Kuramochi, K.~Ida, N.~Umeda, Ultrasmall radial polarizer
  array based on patterned plasmonic nanoslits, Appl. Phys. Lett. 101 (2012)
  161119.

\bibitem{16aom1209}
S.~Rek\v{s}tyt\.{e}, T.~Jonavicius, D.~Gailevi\v{c}ius, M.~Malinauskas,
  V.~Mizeikis, E.~G. Gamaly, S.~Juodkazis, Nanoscale precision of {3D}
  polymerisation via polarisation control, Adv. Opt. Mat. 4~(8) (2016) 1209 --
  1214.

\end{thebibliography}

\section*{Acknowledgements}

The authors thank the support of the National Key R\&D Program of China (No.2017YFB1104600) and the National Natural Science Foundation of China (NSFC) 61590930, 61522503, 51335008, and 61435005.
SJ is also grateful for the ChangJiang scholar project at Jilin University and partial support via DP170100131 grant. We acknowledge discussion on properties of optical vortex beams with Etienne Brasselet.  
\newpage
\appendix 
\section{Supplement}\label{suppl}
\setcounter{figure}{0}    

\begin{figure}[h]
\centering
\includegraphics[width=0.7\textwidth]{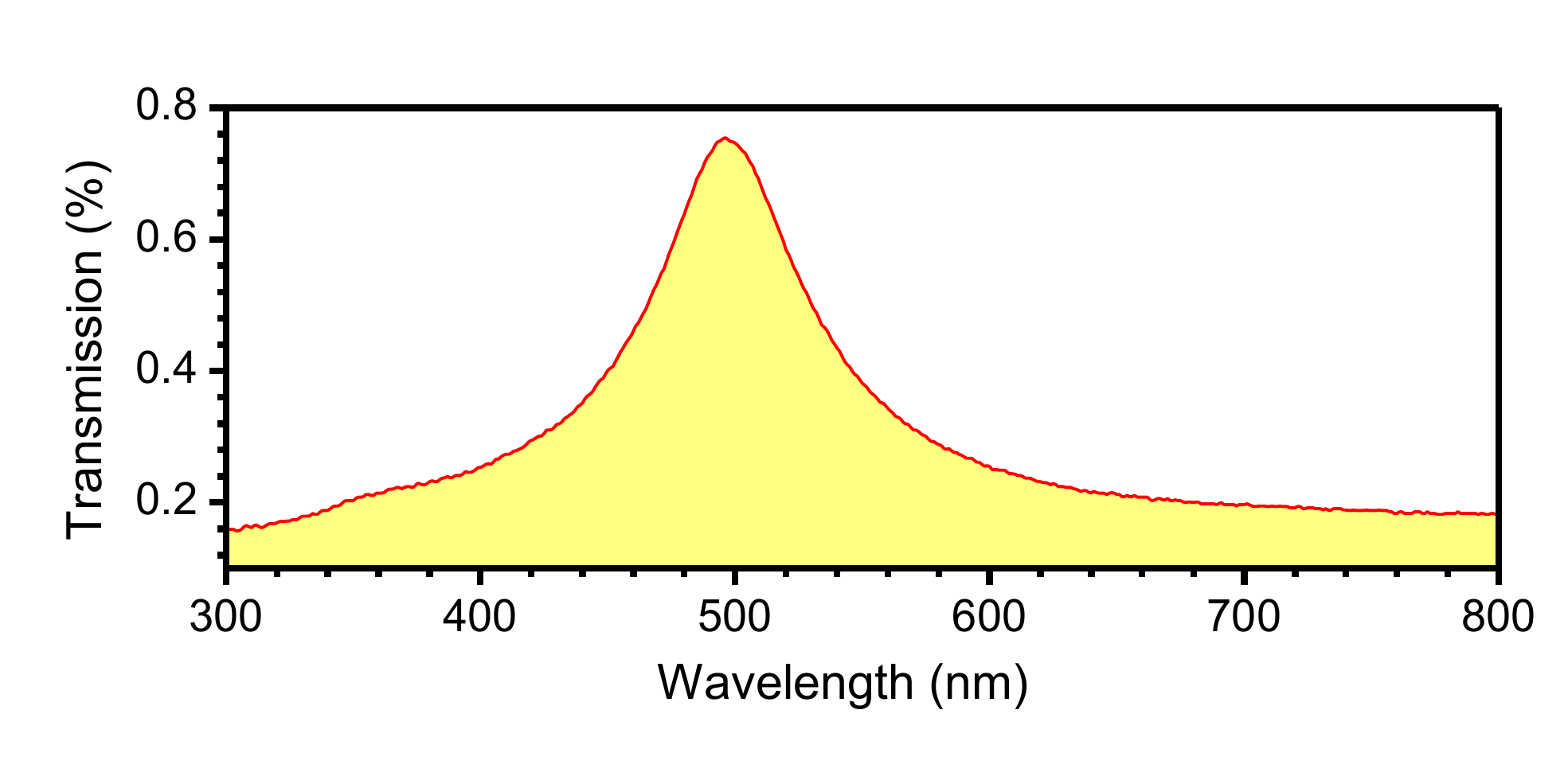}
\caption{Transmission spectrum of the used 50-nm-thick Au film use for laser inscription of the $q$-plates.} \label{f-50nm}
\end{figure}

Figure~\ref{f-50nm} shows transmission spectrum of the unstructured 50-nm-thick gold film on silica. At the used 633~nm wavelength used to characterise $q$-plates transmission was very low 0.22\%. At the Au plasmonic peak of gold $\sim 500$~nm the background transmission can be up to $2.5^\times$ higher, however, even then it is below 1\%.  

\begin{figure}[h]
\centering
\includegraphics[width=1\textwidth]{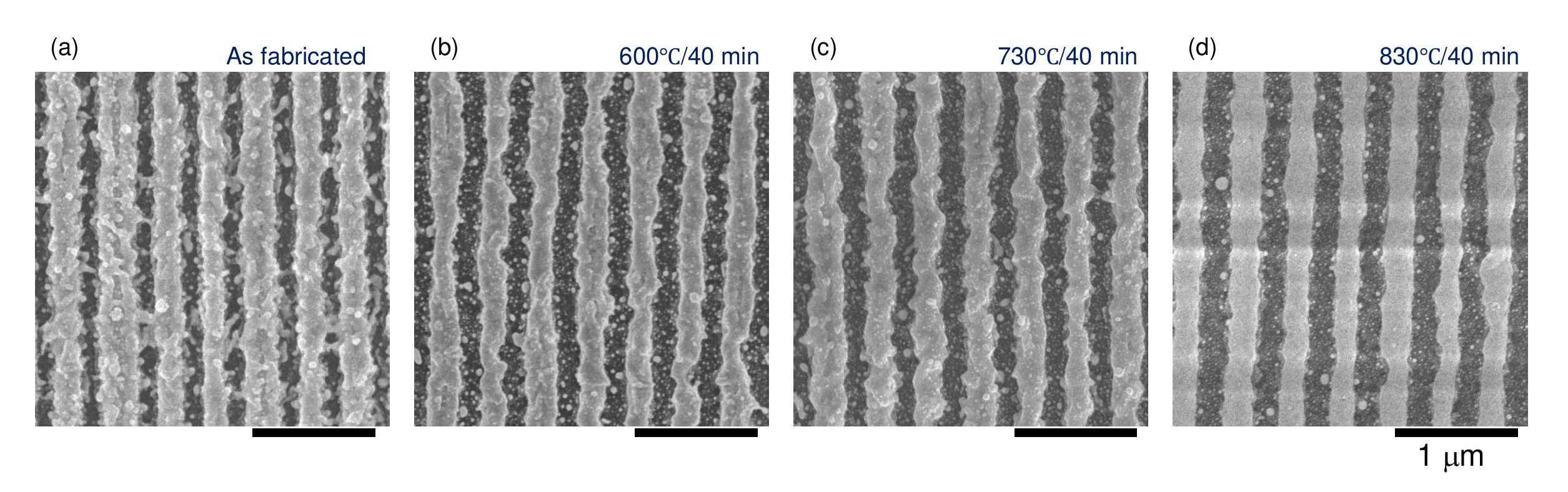}
\caption{Structural changes of Au-on-silica patterns induced by sequential annealing at high temperatures. Ablation was carried out by 343~nm/300~fs pulses.} \label{f-anne}
\end{figure}

Figure~\ref{f-anne} shows structural changes of the gold ablated film after sequential annealing at higher temperatures. 

\begin{figure}[t]
\centering
\includegraphics[width=0.75\textwidth]{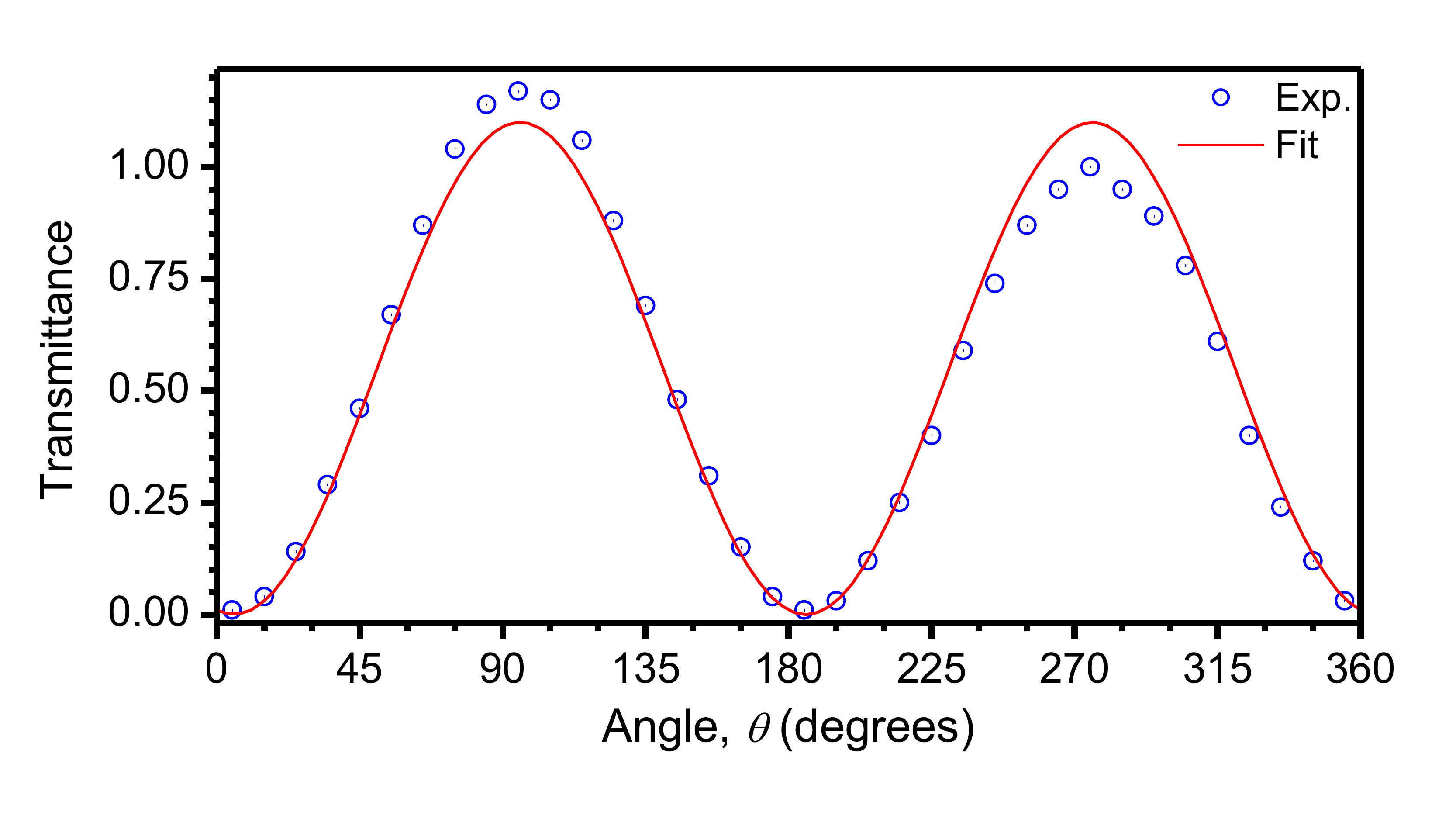}
\caption{Transmission of the linearly polarised 633~nm light through an ablated grating in a Au film ($q=0$ plate of 0.4~mm diameter)  at different angles, $\theta$, of the linear analyser (setup adapted from Fig.~\ref{f-inten}(a)). Fit was made with a function proportional to the Stokes parameter $S_0=\frac{1}{4}[1 + \cos(\Delta^\circ - 2(90^\circ +\theta))]$~\cite{Iwami}. The best fit was achieved for the retardance (in degrees) $\Delta^\circ = (11\pm 2)^\circ$ or 3.1\% of the wavelength, or $\Delta n d = 19.3$~nm. The thickness of Au film was $d = 50$~nm; hence $\Delta n = 0.39$. Refractive index of gold $n+i\kappa =  0.18344 + i3.4332$ at 633~nm. } \label{f-D}
\end{figure}


\end{document}